\documentclass[12pt,preprint]{aastex}

\def\feka{Fe K$\alpha$}     
\def\chandra{{\it Chandra}}     
\def\xmm{{\it XMM-Newton}}     
\def\swift{{\it Swift}}     
\def\asca{{\it ASCA}}     
     
\def\rxte{{\it RXTE}}     
\def\sax{{\it BeppoSAX}}

\def\glast{{\it GLAST}}     
     
\def\lum{erg s$^{-1}$}     
\def\flux{erg cm$^{-2}$ s$^{-1}$}     
\def\nh{cm$^{-2}$}     
\def\arcsec{$^{\prime\prime}$}     
\def\deg{$^{\circ}$}          
     
\def\1136{1136--135}     
\def\1150{1150+497}     
     
\def\ltsima{$\; \buildrel < \over \sim \;$}     
\def\simlt{\lower.5ex\hbox{\ltsima}} 
\def\gtsima{$\; \buildrel > \over \sim \;$}     
\def\simgt{\lower.5ex\hbox{\gtsima}} 

\input{psfig.sty}     
     
\begin{document}     
     
\title{The jet/disk connection in AGN:      
\chandra\ and \xmm\ observations of three powerful radio-loud quasars}

     
\author{Rita M. Sambruna}      
\affil{NASA's Goddard Space Flight Center, Code 661, Greenbelt, MD 20771}     
     
\author{Mario Gliozzi}     
\affil{George Mason University, Dept. of Physics and Astronomy and School of     
Computational Sciences, MS 3F3, 4400 University Drive, Fairfax, VA 22030}     
     
\author{F. Tavecchio and L. Maraschi}     
\affil{INAF/OAB, via Brera 28, 20121 Milano, Italy}     
     
\author{Luigi Foschini}      
\affil{INAF/IASF-Bologna, via Gobetti 101, 40129 Bologna, Italy}     
     
\begin{abstract}      
     
The connection between the accretion process that powers AGN and the
formation of jets is still poorly understood. Here we tackle this
issue using new, deep \chandra\ and \xmm\ observations of the cores of
three powerful radio loud quasars: 1136--135, 1150+497 (\chandra), and
0723+679 (\xmm), in the redshift range $z$=0.3--0.8. These sources are
known from our previous \chandra\ snapshot survey to have kpc-scale
X-ray jets. In 1136--135 and 1150+497, evidence is found for the
presence of diffuse thermal X-ray emission around the cores, on scales
of 40--50 kpc and with luminosity L$_{0.3-2~keV} \sim 10^{43}$ \lum,
suggesting thermal emission from the host galaxy or a galaxy
group. The X-ray continua of the cores in the three sources are
described by an upward-curved (concave) broken power law, with photon
indices $\Gamma_{soft} \sim 1.8-2.1$ and $\Gamma_{hard} \sim 1.7$
below and above $\approx 2$ keV, respectively. There is evidence for
an unresolved \feka\ line with EW $\sim$ 70 eV in the three
quasars. The Spectral Energy Distributions of the sources can be well
described by a mix of jet and disk emission, with the jet dominating
the radio and hard X-rays (via synchrotron and external Compton) and
the disk dominating the optical/UV through soft X-rays. The ratio of
the jet-to-disk powers is $\sim 1$, consistent with those derived for
a number of gamma-ray emitting blazars. This indicates that near
equality of accretion and jet power may be common in powerful
radio-loud AGN.
     
\end{abstract}      
     
     
\section{Introduction}      
     
In recent years, \chandra\ observations of jets hosted by powerful  
quasars with FRII radio morphology revealed that these structures  
transport large ($10^{45-48}$ \lum) amounts of energy as bulk kinetic  
energy from the cores to the distant radio lobes. Only a small  
fraction ($<1$\%) of this energy is expended as radiation at large  
distances (Sambruna et al. 2004), while a larger fraction can be  
released close to the nucleus on sub-pc scales.  
     
In the standard view of Active Galactic Nuclei, the ultimate origin of     
their power is accretion onto a central supermassive black hole (Urry     
\& Padovani 1995). The large energy carried by jets must thus     
originate in the source inner regions, near the black hole. Indeed,     
the formation of the jet should be intimately connected to the accretion     
process, through the spin of the black hole (Blandford \& Znajek 1977)     
and/or centrifugal forces due to the magnetic field threaded by the     
disk (Blandford \& Payne 1982).  
     
A clear connection between the jet and the accretion disk was inferred
by observations of Galactic microquasars and binaries, where the
ejection of a radio component from the center is usually associated to
a minimum of the high-energy flux, and presumably to the disappearance
of plasma in the inner disk orbits (e.g., Fender \& Belloni 2004; but
see Corbel et al. 2004 for a case of a jet in a high/soft state). In
AGN, a correlation between superluminal radio ejecta and X-ray
variability from the nucleus was directly observed in 3C~120 (Marscher
et al. 2002). However, in most cases the notion of a jet/accretion
link so far relied on indirect evidence.  A proportionality between
the kinetic power carried by the jet derived from the lobe properties
and the accretion luminosity, estimated from the O[III] optical lines,
was first inferred by Rawlings \& Saunders (1991) and later confirmed
by Xu, Livio, \& Baum (1999).  A similar correlation using the broad
emission lines as a measure of the accretion power and VLBI data for
the jet was derived by Celotti, Padovani, \& Ghisellini (1997). From
the modeling of the radio-to-gamma-ray Spectral Energy Distributions
of a number of blazars\footnote[1]{Blazars are defined as radio-loud
AGN whose emission is dominated by non-thermal radiation from a
relativistic jet oriented close to the line of sight. Blazars include
BL Lacertae objects, Flat Spectrum Radio Quasars, and Optically
Violently Variable quasars.} with broad emission lines and adequate
data at high energies ($>10$ keV), Maraschi \& Tavecchio (2003) found
that the jet and accretion powers in powerful blazars are of the same
order, suggesting an important role of the black hole rotation to
supply the necessary jet power.
     
In order to estimate the jet and disk powers separately it is
necessary to use objects in which there is {\it direct} evidence for
the presence of an accretion disk as well as of a bright jet. Such
objects are not plentiful as the jet emission is bright only for
orientations close to the line of sight, and in this case it dilutes
disk-related features. Moreover, high energy data are important to
constrain models. In general Flat Spectrum Radio Quasars (FSRQs) meet
these requirements because their non-thermal jet continuum peaks at
IR/optical wavelengths (Sambruna et al. 1996; Fossati et al. 1998) and
thus may yield negligible contribution to the optical/UV region where
the disk/line emission is expected.  So far, such thermal contribution
has been detected in a handful of cases, 3C~273 (von Montigny et
al. 1997), 3C~345 (Bregman et al. 1986), and SWIFT~J0746.3+2548, a $z$=2.979 blazar
recently discovered with \swift\ at energies $>$ 15~keV (Sambruna et
al. 2006a).
     
In this paper, we discuss the jet/accretion connection using
\chandra\ and \xmm\ observations of the cores of three quasars with   
powerful radio emission and FRII morphology: 0723+679 ($z$=0.846),  
1136--135 ($z$=0.554), and 1150+479 ($z$=0.334).  The three sources  
were found to exhibit one-sided X-ray jets on kpc scales from our  
\chandra\ exploratory survey (Sambruna et al. 2002,  
2004). Interestingly, an Fe~K emission line at 6--7~keV (rest-frame)
was detected in the short ACIS-S spectra of the cores of 1150+497 and
0723+679 (Gambill et al. 2003). The Fe line detection indicates that
even in the X-ray band the beamed jet emission does not completely
swamp the accretion-related emission (Grandi \& Palumbo 2004),
qualifying the three sources as excellent candidates to investigate the
jet/disk connection. Moreover, for these objects the jet multifrequency
emission can be measured both close to the core and at large distances
(kpc scales), probing the power transported along the jet.
     
We later acquired longer \chandra\ observations of 1136--135 and     
1150+497 to study their jets, and an \xmm\ exposure of 0723+679 to     
confirm the Fe-K line. The jet properties from the ACIS deep images of     
1136--135 and 1150+497 were discussed in Sambruna et al. (2006b; S06 in     
the following) and Tavecchio et al. (2006). Here we report about the     
core X-ray properties of the three sources. Their basic properties and
radio classifications are listed in Table~1. 
     
The structure of the paper is as follows. In \S~2 we summarize the     
previous \chandra\ observations of the cores. The description of the     
new X-ray observations, data reduction, and analysis protocol is given     
in \S~3. Results from a spatial, timing, and spectral analysis are     
presented in \S~4, while Discussion and Conclusions follow in     
\S~5. Throughout this work, a concordance cosmology with H$_0=71$ km     
s$^{-1}$ Mpc$^{-1}$, $\Omega_{\Lambda}$=0.73, and $\Omega_m$=0.27     
(Spergel et al. 2003) is adopted. With this choice, 1\arcsec\     
corresponds to 7.7 kpc for 0723+679, 6.4~kpc for 1136--135, and     
4.8~kpc for 1150+497. The energy index $\alpha$ is defined such that     
$F_{\nu} \propto \nu^{-\alpha}$.     
     
\section{Observations and Data Reduction}       

\subsection{Deep \chandra\ observations of 1136--135 and 1150+497}      
     
Follow-up \chandra\ observations were carried out on April 16, 2003
for 1136--135 and on July 18, 2003 for 1150+497, with total exposures
81~ks and 70~ks, respectively. Screening was performed using standard
criteria (S06). After screening, the effective exposure times are
70.2~ks for 1136--135 and 61.7~ks for 1150+497.  The net count rates
of the cores in the energy range 0.3--10 keV from an extraction circle
with radius 1.5\arcsec\ are 0.27$\pm$0.02 c/s for 1136--135 and
0.55$\pm$0.03 c/s for 1150+497. Figure~1 shows the 0.3--10~keV images
of both sources (S06).
     
Since we expected bright X-ray cores, we used $\frac{1}{8}$ subarray
mode to reduce the effect of pileup of the nucleus, with an effective
frame time of 0.44~s, or the minimum allowed. With this precaution,
the core of 1136--135 had 0.61 counts/frame, corresponding to
negligible pileup (4\%); however, the core of 1150+497 still had 1.25
counts/frame, or 11\% pileup.
     
We inspected the ACIS-S background for flares and found none.     
Background spectra and light curves were extracted from source-free     
regions on the same chip of the source.  Two sets of spectra were     
extracted. The first set was derived from a circle centered on the     
source X-ray centroid with radius 1.5\arcsec.  As these spectra are     
affected by pileup, during the fit the spectral component     
\verb+pileup+ model in \verb+XSPEC+ was included. The second     
set of spectra were extracted from an annulus, centered on the source     
X-ray position and with inner and outer radii 1\arcsec\ and 3\arcsec,     
respectively. The annuli spectra are mainly contributed to by the     
PSF's wings, which are not affected by pileup at the counts/frame of     
our sources; however, they have a low signal-to-noise ratio.      
     
To assess the reliability of the \verb+pileup+ correction for the     
circle spectra, we compared the spectral fits to the latter with the     
spectral fits to the annuli spectra, using the same continuum models     
(see below). We found consistency of fitted parameters within the     
uncertainties. Thus, in the following we will discuss the circle     
spectra, which have a higher signal-to-noise ratio.  The spectral      
response files were constructed using the corresponding thread in     
\verb+CIAO+ 3.1.      
     
\subsection{\xmm\ observations of 0723+679}      
     
We observed 0723+679 with \xmm\ on April 11, 2005 for a total exposure of     
46.9~ks with the EPIC pn, and 46~ks with the MOS. The count rate of     
the source in 0.3--10 keV is 0.378 $\pm$ 0.005 c/s with the pn and     
0.101 $\pm$ 0.002 c/s with the MOS. As not enough counts were collected     
with the RGS ($<$ 300) for a detailed analysis, here we concentrate on     
the EPIC and OM data only. The MOS1 image in 0.3--10~keV is shown in
Figure~1. 
     
All the EPIC cameras were operated in full-frame mode with a thin     
filter. The recorded events were screened to remove known hot pixels     
and other data flagged as bad; only data with \verb+FLAG+=0 were used.     
The data were processed using the latest CCD gain values, and only     
events corresponding to pattern 0--12 (singles, doubles, triples, and     
quadruples) in the MOS cameras and 0--4 (singles and doubles only,     
since the pn pixels are larger) in the pn camera were accepted.  The     
RGS data do not have sufficient signal-to-noise ratio to perform a     
meaningful analysis.      
     
Unfortunately, our observations were plagued by several episodes of     
intense background flares. For the flare rejection, we considered cuts     
on the count rate of the total hard X-ray (E$>10$ keV) background     
light curve ranging between 1 and 8 c/s for the EPIC pn, and between     
0.35 and 5 c/s for the MOS cameras.  The choice of the count rate     
threshold represents a tradeoff between the necessity to minimize the     
contribution of the flaring background and the need to retain     
sufficient photon statistics for an accurate spectral analysis.  We     
found that for the EPIC pn a cut at 5 c/s is a conservative but still     
suitable background threshold, whereas 2 c/s is a reasonably good     
choice for the MOS cameras. Excluding these events reduces the     
effective total exposures times to 19.9~ks for the EPIC pn and 35.2 ks     
for the MOS cameras. There is no pile-up in the pn or MOS cameras     
according to the {\tt SAS} task {\tt epatplot}.     
     
The EPIC pn and MOS spectra were extracted from a circular region
centered on the source's position and with radius 32\arcsec. The
background was extracted from a circle with radius 60\arcsec,
positioned in a region of the CCD free from serendipitous X-ray
sources. Response files were created with \verb+SAS+ v.6.1. There is
no pile-up in the pn or MOS cameras according to the SAS task {\tt
epatplot}.
     
The data from the Optical Monitor (Mason et al. 2001) were processed
with the latest (December 3, 2005) calibration files. The
observed magnitudes, extracted from the output files of the pipeline
and corrected for systematics, are: $UVM2=17.1\pm 0.1$
($\lambda=2310$~\AA), $UVW1=17.8\pm 0.1$ ($\lambda=2910$~\AA),
$U=18.4\pm 0.1$ ($\lambda=3440$~\AA), and $B=19.1\pm 0.1$
($\lambda=4500$~\AA). These values were corrected for the Galactic
absorption column ($A_{\rm V}=0.132$) and converted into fluxes by
using standard formulae (e.g., Zombeck 1990).
     
\section{Results}      
     
\subsection{Chandra spatial analysis}      
     
Inspection of the 0.3--10 keV images of 1136--135 and 1150+497 in
Figure~1 reveals the presence of faint diffuse emission around the
cores of both targets. To quantify this result we performed a detailed
spatial analysis. The following procedure was adopted. First, radial
surface-brightness profiles were extracted from a series of concentric
annuli centered on the radio core position. Off-nucleus X-ray point
sources, as well as the X-ray jet, were excluded. Second, the radial
profiles were fitted with a model including the instrument Point
Spread Function (PSF). The latter was created using the Chandra Ray
Tracer (\verb+ChaRT+) simulator which takes into account the spectral
distribution of the source. We used the best-fit X-ray continua from
the spectral analysis (Table~2).  The significance of adding the PSF
model was determined using an F-test, assuming as threshold for
significant detection a probability P$_F=$99\%.

The observed radial profiles of the two sources in the total energy     
band 0.3--10 keV are shown in Figure~2. Comparing the instrumental PSF     
(dashed line) with the data, excess X-ray flux over the PSF's wings is     
apparent above 5--6\arcsec, indicating the presence of diffuse     
emission around the cores. To model this component, we used a $\beta$     
model, described by the following formula (e.g., Cavaliere \&     
Fusco-Femiano 1976):     
     
\begin{displaymath}     
S(r)=S_0\left(1+{r^2\over r_c^2}\right)^{-3\beta+1/2},      
\end{displaymath}     
     
where $r_c$ is the core radius. The radial profiles were then fitted     
with a model including the PSF, the background (held fixed at the     
measured value, $4.1 \times 10^{-6}$ c/s/arcsec$^2$), and the     
$\beta$-model. The latter is required at P$_F >$ 99\% confidence in the     
fits for both sources.  The fitted parameters are: $S_0=     
(2.2\pm1.0)\times10^{-5}{\rm ~ct~s^{-1}~arcsec^{-2}}$, $\beta=0.59     
\pm 0.15$, $r_c=(7.3 \pm 4.4)$\arcsec, or $\sim$ 47 kpc for 1136--135;      
$S_0= (4.1\pm1.1)\times10^{-5}{\rm ~ct~s^{-1}~arcsec^{-2}}$,
$\beta=0.64 \pm 0.07$, $r_c=(7.5 \pm 1.1)$\arcsec, or $\sim$ 36 kpc
for 1150+497. The core radius is on the low-end of the range observed for
intermediate-$z$ quasars (Crawford \& Fabian 2003), and suggests
emission on the scale of the host galaxy or a group of galaxies. The
best-fit $\beta$ models are plotted in Figure~2 (dot-dashed
lines). The bottom panels show the residuals of the fits to the radial
profiles.
     
To analyze the 0.5--8~keV spectrum of the diffuse emission we
extracted the count rate from the diffuse emission in 1136--135 from
an annular region of inner and outer radii 5\arcsec\ and 15\arcsec, 
with the jet excised. The inner radius of the extraction region is the
minimum distance from the core at which the extended emission starts
exceeding the PSF radial profile, whereas the outer radius represents
the distance at which the background component dominates the total
emission.  The total count rates of the diffuse emission in this
annulus are $(4.1\pm0.7)\times 10^{-3}$ c/s in the case of 1136--135
and $(8.6\pm0.8)\times 10^{-3}$ for 1150+497. 
     
The spectra of the diffuse emission in both sources were fitted with a
model including a thermal (\verb+apec+) component and a power law, the
latter accounting for the contribution of the wings of bright core
PSF\footnote[1]{While the best-fit to the core X-ray spectra is more
complex than a single power law, we find that due to the low
signal-to-noise ratio of the diffuse emission spectra the best results
are obtained by using a power law.}, with Galactic absorption acting
on both components. The photon index of the power-law component was
fixed at the best-fit value obtained in the fit of the core spectrum
($\Gamma$=1.80 and 1.91, for 1136--135 and 1150+497); the temperature
was constrained to vary in the range 0.1 and 1.5 keV, typical of the
host galaxy halo. Since the metal abundance $A$ relative to solar is
poorly determined when left free to vary, we adopted the following
procedure: We fixed $A$ to 0.2, 0.4, 0.6, 0.8, 1 and compared the
respective fits. For 1136--135, $kT$ is relatively well determined for
$A=0.2-0.6$; for 1150+497, no significant differences in the spectral
fits are found when $A$ spans the entire 0.2-1 range.  For $A=0.6$,
the measured temperature and 90\% uncertainties is
$kT=0.15^{+0.3}_{-0.05}$ keV for 1136--135 and $kT=0.1^{+1.0}_{-0.0}$
keV for 1150+497. 

The observed fluxes of the thermal component in 0.3--2 keV are     
F$_{0.3-2~keV} \sim 4.1 \times 10^{-15}$ \flux\ for 1136--135, and     
F$_{0.3-2~keV} \sim 2.0 \times 10^{-14}$ \flux\ for 1150+497. These     
are consistent with the upper limits derived from our previous short     
exposures (Gambill et al. 2003). The corresponding intrinsic     
luminosities are L$_{0.3-2~keV} \sim 7.5 \times 10^{42}$ \lum\ and      
L$_{0.3-2~keV} \sim 9.0 \times 10^{42}$ \lum, respectively.      
     
We are interested in deriving the density of the hot gas around the     
cores of 1136--135 and 1150+497, an important parameter in our model     
of jet deceleration through gas entrainment (Tavecchio et     
al. 2005). The gas density can be obtained by de-projecting the     
surface-brightness profiles (see, e.g., Ettori 2000):     
     
\begin{equation}     
n(r)=n_0\left(1+{r^2\over r_c^2}\right)^{-{3\beta\over 2}}     
\end{equation}     
     
This relation assumes isothermal, hydrostatic equilibrium in spherical
symmetry.  We adopted the cooling function value (Sarazin 1988) for
the hot gas temperature and an abundance $A$=0.6. The resulting
central particle densities for 1136--135 and 1150+197 are
$n_0=(5.1\pm2.3)\times 10^{-4}~\rm{cm}^{-3}$ and
$n_0=(7.5\pm2.0)\times 10^{-4}~\rm{cm}^{-3}$ for 1136--135 and
1150+497, respectively.
     
We evaluated the relative contribution of the thermal diffuse
component and the point source in the region of extraction of the core
spectra, a circle of radius 1.5\arcsec. The ratios between the
integrated $\beta$ and PSF model emission are 2.3$\times 10^{-4}$ for
1136--135 and 1.9$\times 10^{-4}$ for 1150+497. These estimates are,
however, affected by large uncertainties ($\simeq 40-50\%$), due to
the large errors on the spatial parameters. Taken at face value, these
ratios imply that the diffuse emission gives a negligible contribution
($<$ 0.1\%) to the total emission within 1.5\arcsec.
     
In summary, we find that the cores and jets of 1136--135 and 1150+497     
are embedded in soft ($kT \sim 0.1$ keV) thermal X-ray emission on a     
scale of tens of kpc and with luminosities $\sim 10^{42}-10^{43}$     
\lum. The physical scale of the thermal diffuse emission     
is several tenths of kpc, likely related to the halo of the host
galaxy or a small group of galaxies. The presence of the ISM and its
properties impact discussions of mechanisms for jet deceleration on
kpc scales (Tavecchio et al. 2006 and references therein).
     
\subsection{Timing analysis}      
     
We searched for X-ray flux variability in the background-subtracted
light curves of the cores. No timing analysis was performed for the
core of 0723+679, because of the numerous large-amplitude flares
affecting the background light curve.

The light curves of 1136--135 and 1150+497 are shown in Figure~3.
According to the $\chi^{2}$ test, no significant variability of the
0.3--10 keV flux is detected in 1136--135, with a constancy
probability P$_{const} \sim 90\%$. On the contrary, in 1150+497 there
is an indication that the X-ray flux is variable, with P$_{const}
<1\%$. From Figure~3, the X-ray flux increases monotonically toward
the end of the observation by a factor 1.2 in a few hours. 

To investigate the energy-dependence of variability of 1150+497, light
curves were extracted in the spectral bands 0.3--1.5~keV and
1.5--10~keV. The choice of these energy ranges is motivated by the
results of the continuum spectral analysis, which shows a break around
1.5~keV of the best-fit broken power law model (Table~2a). With this
division, roughly equal counts are contained in the two energy ranges,
$\sim$ 0.20 c/s at soft and 0.35~c/s at hard energies. We find
evidence of variability at both soft and hard X-rays, with P$_{const}
\sim 1.3\%$ and 8.9\%, respectively. The fractional variability
amplitude (Gliozzi et al. 2003) in the two energy ranges is
$F_{var_soft}=0.034 \pm 0.009, F_{var_hard}=0.043 \pm 0.013$. Thus,
there is no significant dependence of the amplitude of variability on
energy.
      
The hardness ratios, defined as the ratio of the 1.5--10~keV counts to
the 0.3--1.5~keV counts, are also plotted in Figure~3. Neither
1150+497 nor 1136--135 show any indication of spectral variability,
with P$_{const} \sim 75-80\%$ for both sources.  

\subsection{Spectral analysis}      
     
The ACIS and EPIC spectra, extracted as described in \S~3, were
grouped so that each new bin had at least \gtsima 20 counts to enable
the use of the $\chi^{2}$ statistics. Spectral fits were performed
within \verb+XSPEC+ v.11.2.0 on the energy ranges 0.5--8 keV for ACIS,
0.3--10 keV for EPIC pn, and 0.4--10 keV for EPIC MOS, where the
calibrations for spectral analysis are best known and the background
negligible.
     
In all fits, an absorption column density fixed to the Galactic value
was included. In the case of 1136--135 and 1150+497, where modest
amounts of pileup is present, the models used to fit the ACIS spectra
included a component correcting for pileup (model \verb+pileup+ in
\verb+XSPEC+). In this model, the timeframe was fixed at 0.441~s and     
the \verb+psffrac+ parameter at 0.95.     
     
The spectral results are summarized in Table~2 (X-ray continua) and     
Table~3 (the Fe-K line). All uncertainties are 90\% confidence for one     
parameter of interest ($\Delta\chi^2$=2.7). Below we comment     
separately on the spectral fit results for the continuum and the Fe     
line.      
     
\subsubsection{The Continuum}     
     
The X-ray spectra of the cores were fitted at first with a single
power law model and Galactic column density (Table~1). This provides a
formally acceptable fit to the EPIC spectrum of 0723+679 and to the
ACIS-S spectrum of 1136--135, but not to 1150+497. Inspection of the
residuals shows the presence of excess flux below 1 keV, indicating
the presence of a soft X-ray spectral component in all three
sources. We know that, at least in the case of 1136--135 and 1150+497,
the diffuse emission does not contribute significantly in the
extraction region of the core spectrum (\S~3.1); for 0723+679 the
addition of a thermal component (\verb+apec+) to the power law
improves the fit significantly, however, the fitted temperature is
0.08 keV (for a fixed abundance of 0.2 solar), effectively mimicking a
steep power law at low energies. We thus conclude that the soft X-ray
excess flux below 1 keV is intrinsic to the point source in all three
cases. 
          
Among the phenomenological models, a broken power law describes the X-ray     
spectra best. In all three sources, the fitted low-energy slope is     
steeper than the higher-energy slope, thus yielding a concave (upward     
curved) continuum. The parameters of the broken power law best-fit     
models are listed in Table~3, while Figure~4 shows the EPIC and     
ACIS-S residuals of the broken power law model fits.      
   
We can not distinguish between a non-thermal and a thermal origin for
the soft excess in the ACIS and EPIC data. We fitted the data with a
power law at hard energies plus a blackbody to model the soft X-rays,
in the assumption that the soft X-ray emission originates from the
accretion disk. While the addition of the blackbody improves the fit
significantly (P$_F >$ 99\%), the derived temperatures are \gtsima 1
keV, too large to be related to emission from a Shakura-Sunyaev
accretion disk; the latter is expected to peak in the EUV portion of
the spectrum, implying $kT \approx$ 0.1~keV. Clearly, such a model
would produce a steep power law-like continuum in the soft part of the
ACIS-S and EPIC energy range. 
  
Prompted by the detection of an \feka\ line in 1136--135 and 1150+497,
we tried more complex fits to the ACIS-S continua of these sources
using models including reflection from a neutral absorber
(\verb+pexrav+). Reflection from a neutral medium provides a formally
good fit to the data of both 1136--135 and 1150+497, with reduced
$\chi^2$=1.03 and 1.15, respectively. However, the fitted reflection
fraction, $R \sim 2.5-3$, is much larger than the value expected from
the line Equivalent Widths (EW) according to the relation
EW=160$\times \Omega/2\pi$, or $R$=1.25 (George \& Fabian
1991). Fixing the reflection fraction to the latter value gives an
acceptable fit, but not statistically better than a broken power law.
We conclude that reflection from a cold medium, while consistent with
the data, is not required.
     
\subsubsection{The \feka\ emission line}      
      
As apparent from Figure~4, line-like residuals are present around
4--5~keV (6--7~keV rest-frame) in 1136--135 and 1150+497. Thus, we
fitted the ACIS-S data adding a Gaussian component to the best-fit
continuum model. The Gaussian gave a significant $\chi^2$ improvement
only for 1136--135 and 1150+497, with $\Delta\chi^2=6$ for 2
additional parameters, corresponding to P$_F$=98\% and 96\% for
1136--135 and 1150+497, respectively. A more conservative test using
Monte Carlo simulations (Protassov et al. 2002) yields a significance
for the lines of 96\% and 94\%.
     
The confidence contours of the Gaussian width versus its energy show     
that the Fe line is unresolved in both sources, and consistent with     
both a narrow unresolved ($\sigma=0.01$ keV) and broad Gaussian     
line. Thus, we report results for both a narrow ($\sigma \equiv 0.01$     
keV) and a broad ($\sigma \equiv 0.25$ keV) Gaussian. The fitted     
parameters are listed in Table~3. The rest-frame energy of the line     
for both 1136--135 and 1150+497 is consistent with the \feka\     
fluorescent line often detected in lower-luminosity sources. The     
measured Equivalent Widths are in the range 60--80 eV.     
     
In the case of 1136--135 the Fe-K residuals appear more complex than a
single Gaussian. Indeed, adding a second narrow Gaussian component to
the brokwn powerlaw plus narrow line model yields a significant
improvement of the fit, $\Delta\chi^2=7.4$ for 2 additional dofs,
significant at P$_F$=97.5\%. The fitted rest-frame energy of the
second line is $6.8 \pm 0.1$ keV and its EW $\sim 50$ eV. The EW of
the 6.4~keV line does not change. Taking the second line energy at
face value, the closest emission line would be FeK$\beta$ at 7.06~keV,
or K$\alpha$ from H- or He-like Fe. This finding indicates additional
complexity in the X-ray spectrum of 1136--135 which warrants future
investigation with higher-quality X-ray data.
     
In the EPIC data of 0723+679 there is no strong evidence for
significant line-like residuals at the higher energies (Fig.~4),
contrary to the earlier \chandra\ spectrum (Gambill et
al. 2003). Adding a narrow Gaussian to the best-fit continuum model
yields only a modest improvement of $\Delta\chi^2=3$ for 2 additional
parameters (the energy line was fixed at 6.4 keV in the source
rest-frame), corresponding to an 80\% significance from both the
F-test and the Monte Carlo simulations. However, the line is detected
(i.e., its flux is different from zero) at 90\% confidence. The lack
of a stronger detection is not surprising, considering the substantial
loss of data due to the flaring background. The 99\% upper limit to
the line Equivalent Width is EW $<$ 100 eV, fully consistent with the
\chandra\ detection. 
     
In summary, we confirm the detection of an Fe-K line in 1150+497 and     
report its first detection in 1136--134. The \feka\ line is detected     
but not required in 0723+679, with a 99\% confidence upper limit to     
its EW of 100 eV.     
  
\section{Discussion}      

\subsection{The SEDs: a mix of jet and accretion disk emission}     

An outstanding issue in the study of radio-loud sources is the
relative contribution of the accretion disk and of the unresolved jet
to the emission from the core. To this end, the detection of the
\feka\ emission line in the X-ray spectrum, together with the
knowledge of the jet parameters, can provide important clues.

\subsubsection{The \feka\ line: dilution by beaming?}       

Contrary to radio-quiet sources the presence of the \feka\ line
emission from the cores of radio-loud AGN is not well
established. Previous \asca, \rxte, and \sax\ observations detected Fe
K-shell emission from the cores of the brightest Broad-Line Radio
Galaxies (e.g., Sambruna, Eracleous, \& Mushotzky 2002 for a review of
earlier results), and in a handful of radio-loud quasars (Hasenkopf,
Sambruna, \& Eracleous 2000; Reeves \& Turner 2000). It is generally
found that the \feka\ line and associated reflection continuum in
BLRGs are much weaker than in their RQ counterparts, the Seyfert~1s
(e.g., Sambruna et al. 2002). A recent EPIC observation of the BLRG
3C~111 also shows a weak (EW $\sim$ 60 eV) but resolved \feka\ emission
line (Lewis et al. 2005).
     
The simplest interpretation is that RL sources have standard disks and     
the reflection features are diluted by the beamed emission of the     
jet. Alternatively, narrow lines could come from the outer portions of     
an ADAF (Eracleous et al. 2000). However, it has proven difficult to     
disentangle the jet and isotropic contributions in BLRGs due to the     
lower jet/disk luminosity ratios in these sources.    
      
The targets of this paper are higher-luminosity quasars (Table~1) for     
which the presence of beamed jet emission can be inferred from the 
large-scale jet properties. The presence of the accretion-related Fe line     
from their cores suggests that the beamed jet component does not fully     
dominate the X-ray emission from the cores. 

The \chandra\ spectra of 1136--135 and 1150+497 are consistent with
both a narrow and a broad \feka\ line. Proposed sites for the origin
of a narrow line, at least in radio-quiet AGN, include the edge of the
pc-scale torus, the Broad-Line Region (BLR), or scattering by gas or
dust along the line of sight. It is conceivable that the observed Fe
lines in Seyferts are a mix of a broad component from the disk and a
narrow core from either/all the locations mentioned above (Nandra
2006). In the case of 1136--135 and 1150+497, where the viewing angle
to the central black hole is small, an origin from the edge of the
torus seems improbable. Similarly, scattering would produce emission
lines at softer X-rays, which are not observed in the ACIS and EPIC
spectra of our sources. While a narrow line from the BLRs cannot be
excluded, the presence of a broad Fe-line from the accretion disk
appears likely. 

Support for this comes from the Eddington ratios of the three quasars.
From Table~1 the Eddington luminosities are L$_{Edd}=6, 6, 4.3 \times
10^{46}$ \lum\ in 0723+679, 1136--135, and 1150+497, respectively. An
estimate of the bolometric (accretion-related) luminosity comes from
the optical/UV energy range of the SEDs, which we argue below is
dominated by the disk emission (\S~4.1.2). The disk luminosities,
L$_{disk}$, are listed in Table~4. Comparing L$_{Edd}$ and L$_{disk}$,
and if the disk luminosity is representative of the bolometric
luminosity, the three quasars are radiating at or very close to the
Eddington limit. A standard accretion disk is thus most likely present
in both powerful quasars. We thus discuss the detected \feka\ line in
1136--135 and 1150+497 assuming that the line is broad, and originates
from the inner radii of a standard accretion disk.

In this scenario, we can estimate the expected EW of the \feka\
emission line for a given inclination angle and photon index (George
\& Fabian 1991). Assuming an inclination angle 4\deg, as derived from
the SED modeling (Table~4), and a photon index $\Gamma=1.9$ typical of
the X-ray emission from radio-quiet AGN, from Figure~14 of George \&
Fabian (1991) the expected EW of the \feka\ line is 140~eV. An
\verb+XSPEC+ simulation including a disk component (power law with
$\Gamma_{disk}=1.9$) plus a Gaussian line with EW=140~eV, plus a jet
component (power law with $\Gamma_{jet}=1.7$), normalized such that
the line EW=70~eV, shows that the jet would be responsible for $\sim$
50\% of the total flux in 2--10~keV. 

We thus conclude that both jet and disk emission contribute to the
X-ray emission from the three quasar cores. Specifically, the beamed
emission contributes roughly 50\% to the total flux in 2--10~keV,
while the disk emission dominates below 2~keV. 

Our results are similar to those obtained for 3C~273 by Grandi \&
Palumbo (2004). A re-analysis of multi-epoch \sax\ spectra of 3C~273
indicated that the jet dominates over the Seyfert-like component above
2 keV, while it is the only source of X-ray emission above 40~keV,
where the reflection component of the Seyfert-like emission cuts off
(Grandi \& Palumbo 2004; Foschini et al. 2006; Turler et
al. 2006). Note that, like 3C~273, the sources in this paper are
characterized by non-negligible beaming, as shown by their one-sided
\chandra\ jets. In this case, dilution of the disk-related features in
the X-ray spectrum is expected. Whether beamed radiation dominates the
X-ray emission of less beamed, lower-Eddington ratio sources, such as
some Broad-Line Radio Galaxies (e.g., Eracleous et al. 2000), remains
to be demonstrated.

\subsubsection{SED modeling}
  
Figure~5 shows the SEDs of the cores of 0723+679, 1136--135, and
1150+497 obtained combining the X-ray best-fit continua from our
\chandra\ observations with literature data at other wavelengths.
The UV data were taken from Kuraszkiewicz et al. (2002), Lanzetta et
al. (1993); the optical data from Gambill et al. (2003) and references
therein; the radio from Condon et al. (1998), K\"uhr et al. (1981),
Becker, White, \& Edwards (1991), Pauliny-Toth et al. (1978). The IR
data for 1150+497 were taken from the 2MASS database and from Impey \&
Neugebauer (1988). Only in the case of 0723+679 the optical-UV fluxes
were derived from simultaneous data taken with the XMM Optical
Monitor.  Other data are not simultaneous or even close in time; we
are not aware of multiwavelength campaigns to study these
sources. None of these sources was detected by EGRET, therefore only
upper limits to the GeV flux are given.

As discussed in \S~4.1.1, there is evidence that both the disk and the
jet contributes to the X-ray emission, with the jet becoming prominent
at harder energies and the disk dominating below 2~keV. At longer
wavelengths, the extremely hard UV spectrum of 1136--135 derived by
Kuraszkiewicz et al. (2002) and the OM data of 0723+679 strongly
suggest that the UV-optical emission in these two sources is dominated
by the UV bump component associated to the disk. The optical and UV
data for 1150+497 (although not simultaneous) support the same
conclusion. On the other hands, the data in Figure~5 indicates that
the radio through IR emission belongs to a different spectral
component than optical/UV, most likely to the jet. We therefore
interpret the observed SEDs using a model that combines the broad-band
jet emission usually considered for blazars and a standard disk
spectrum typical of radio-quiet QSOs.

We parameterize the disk emission in terms of a blackbody (optical to
soft X-rays) plus a flat spectrum reproducing the reflection component
at harder X-rays (Korista 1997).  We can fix the disk luminosity in
all the three sources at the level indicated by the observed
optical-UV continuum. Interestingly, the direct observation of the
disk is usually prevented in blazars by the intense non-thermal jet
emission. In that case the disk luminosity is usually estimated
assuming that the Broad Line Region reprocesses a fraction $f\simeq
0.1$ of the disk continuum, $L_{BLR}=f L_{disk}$ (Maraschi \&
Tavecchio 2003). In the sources under consideration here, the disk and
the BLR luminosities are simultaneously available, allowing to check
the assumption above.  Indeed, using the values of $L_{BLR}$ in Table
1 and the value of $L_{disk}$ fixed by the SEDs, we derive $f=0.05$
for 0723+679, $f=0.3$ for 1136--135, and $f=0.08$ for 1150+497. These
values are consistent with the assumption $f\simeq 0.1$ within the
(probably large) uncertainties involved in this procedure.

For the jet, we use the model fully described in Maraschi \& Tavecchio
(2003) to calculate the synchrotron (radio-IR) and Inverse Compton
(IC) emission. The IC component includes as seed photons the
synchrotron photons themselves (SSC) plus the external radiation field
(EC). In fact, as apparent from Table~1, all three sources have
luminous BLRs, providing an important source of soft photons.  To
calculate the EC emission it is necessary to provide the energy density
of the BLR radiation: given the luminosity of the BLR (Table 1), it is
possible to derive the energy density of the external radiation field
assuming a distance $R_{BLR}$ for the BLR clouds. $R_{BLR}$ has been
directly estimated for a few sources through the reverberation mapping
technique (Kaspi et al. 2005; Bentz et al. 2006). These studies
suggest a proportionality of $R_{BLR}$ and the luminosity of the
illuminating source. From these relations and the adopted luminosities
of the disk we derived the values of $R_{BLR}$ used as input for the
modeling. Not all the parameters are fully constrained. The size of
the emitting region, usually constrained in blazar modeling by the
minimum variability timescale, is fixed here at $R=10^{16}$ cm, a
typical value for blazars with similar power (e.g., Ghisellini et
al. 1998; Tavecchio et al. 2002).

With these choices we calculated the models shown in Figure~5. The jet
model cannot reproduce the data in the radio band, since the
synchrotron emission is self-absorbed below $\sim 10^{11}$ Hz. As 
usually assumed, the radio component is probably due to the integrated
contribution from the emission of the jet at larger distance from the
core. In the model we assume that the high-energy component has the
largest luminosity compatible with the EGRET upper limit. The main
parameter determining the gamma-ray luminosity is the energy of the
electrons emitting at the peak $\gamma_{\rm peak}$. Increasing
(decreasing) its value causes the gamma-ray luminosity to increase
(decrease) without affecting the rest of the {\it observed} SED. As an
example we show in Figure~6 the SEDs of 0723+679 computed for two
different values of $\gamma_{\rm peak}$ and all the other parameters
unchanged.

The physical parameters derived modeling the jet emission do not
differ from those inferred for other powerful blazars (Tavecchio et
al. 2002). A more prominent disk component in the optical-UV band with
respect to the other powerful blazars is reproduced using a slightly
less intense synchrotron continuum, obtained with a smaller magnetic
field. In this respect, data in the IR-submm range would be crucial,
since they would allow to probe the actual level of the peak of the
synchrotron component.


We conclude that the SEDs of the three quasars are best explained by a
mixed contribution of non-thermal jet emission and emission related to
the accretion disk. The jet emission should dominates the SEDs below
IR and above hard X-ray wavelengths, while the disk contributes mainly
to the optical/UV and soft X-rays. However, we stress that the
modeling of the jet component is based on rather poor data. Indeed,
the only strong observational constraint is offered by the X-ray data;
because of the lack of information in the IR-submm region and the
upper limit in the $\gamma $-ray band, we have a large degree of
freedom in the modeling. Thus, future observations focussed in the IR
region and the possible detection in the MeV-GeV band will provide an
important test of our interpretation.

\subsection{Jet power and accretion}     

Important physical quantities can be inferred from the SED modeling,
in particular the power carried by the jet, $P_{\rm jet}$, and its
radiative luminosity (corrected for beaming), $L_{\rm jet}$.  We
derive $P_{\rm jet}$ assuming a composition of 1 (cold) proton per
electron (e.g., Ghisellini \& Celotti 2002; Maraschi \& Tavecchio
2003). The jet power is then dominated by the bulk kinetic power
associated with the proton component. The derived values are reported
in Table~3. Since the estimated kinetic power depends on the total
number of protons, which is in turn equal to the total number of
electrons, the result depends only on the electrons at low energies
which, for a given energy density of the external radiation field, are
determined by the requirement of reproducing the X-ray
spectrum. Therefore, in spite of the large range in some of the
parameters allowed by the poor sampling of the SED, the derived power
can be considered reasonably constrained.

Another input parameter for deriving $P_{\rm jet}$ and $L_{\rm jet}$
is the bulk Lorentz factor $\Gamma$. Here we assume $\delta=\Gamma$,
implying an observing angle $\theta =1/\delta$. From the values of
$\delta$ from Table~4 the inclination angles are 3.8\deg\ for 0723+679
and 1136--135, and 2.8\deg\ for 1150+497. We recall that, for a given
Doppler factor $\delta$, the observing angle {\it must be} less than
$1/\delta$ (e.g., Tavecchio et al. 2004). 

Interestingly, for these sources we also have independent estimates of
the jet power coming from the modeling of the multiwavelength
large-scale jet emission. We can therefore compare the power of the
jet on two vastly different scales (Tavecchio et al. 2004). The
kpc-scale jet power was estimated by us (Sambruna et al. 2002) and is
$1.2\times 10^{47}$ \lum, $3\times 10^{46}$ \lum, and $1.8\times
10^{47}$ \lum\ for 0723+679, 1136--135, and 1150+497, respectively. A
comparison with the pc-scale values reported in Table 4 shows that the
two powers agree within a factor of 2, confirming previous findings
for a group of gamma-ray blazars (Tavecchio et al. 2004). This
suggests that the power channeled into the jet is stable on timescales
of the order of $10^4-10^5$ yrs, and that a negligible fraction of the
kinetic power of the jet is dissipated along its path from the
innermost regions of the AGN to the 100s-kpc scales.

It is quite interesting to compare the jet power derived above with
that of jet in other blazars.  In Figure~7 the jet power is plotted
against the radiative luminosity of the jet (left panel) and the disk
luminosity (right panel) for the three sources of this paper, together
with the blazars of the sample studied in Maraschi \& Tavecchio
(2003). The powers of the newly discovered MeV blazar
SWIFT~J0746.3+2548 (Sambruna et al. 2006a) at $z \sim 3$ are also
reported.

In Figure~7, the powers of 0723+679, 1136--135, and 1150+497 are
clearly consistent with those of other blazars in both plots. The
oblique lines in the left panel mark different jet radiative
efficiencies, $\eta = L_{\rm jet}/P_{\rm jet}$. It can be seen that
the three sources analized here are characterized by smaller radiative
efficiencies, $\eta = 10^{-2}-10^{-3}$, than the majority of blazar
jets, which instead are located in $\eta =10^{-1}-10^{-2}$.

A possible bias in Figure~7 is related to the fact that the radiative
luminosity of the jets in FSRQs with kinetic power similar to
0723+679, 1136--135, and 1150+497 is dominated by the emission in the
$\gamma$-ray band, and that the modeling of the SEDs (from which the
data in Figure~7 were calculated) are generally done assuming the {\it
average} flux of the detections obtained by EGRET. Therefore, the data
in Figure~7 are representative of states of moderately high activity,
and it is likely that the radiative luminosity {\it averaged} over the
duty cycle of the $\gamma$-ray emission is more similar to that
derived for the three sources analized here.  Future $\gamma$--ray GeV
observations, probing the band where most of the power is released
(see Figs.~5 and 6) and better characterizing the $\gamma$-ray duty
cycle, will allow us to better assess the total radiative output,
greatly improving the estimate of the radiative efficiency estimates.

\section{Summary and Conclusions}     

We presented deep \chandra\ and \xmm\ observations of three radio-loud     
quasars with powerful large-scale X-ray jets. The spatial, timing, and     
spectral properties of the cores were discussed. The results are:     

\begin{itemize}      

\item Soft X-ray (0.3--2~keV) diffuse emission is present in the ACIS     
images around the cores of 1150+497 and 1136--135. The diffuse     
emission scale, 30--40~kpc, and luminosity, L$_{0.3-2~keV}     
\sim 10^{43}$ \lum, suggest thermal emission     
from the medium in the host galaxy and/or a galaxy group. The
contribution of the diffuse component to the core X-ray emission
within 1.5\arcsec\ is negligible.

\item The X-ray continua of the cores are well described in all cases     
by a broken power law model, yielding $\Gamma_{soft} \sim 1.8-2.1$,
$\Gamma_{hard} \sim 1.7$, and E$_{break} \sim 1.5$ keV. 

\item The \feka\ emission line is confirmed in 1136--135 and     
1150+497, with a rest-frame EW $\sim$ 70 eV. The short useful EPIC
exposure of 0723+679 unfortunately hinders confirmation of the \feka\
line, previously detected with \chandra.
     
\item Modest (factor 1.2) flux variations are observed in 1150+497     
at both soft (0.3--1.5~keV) and hard (1.5--10~keV) energies within
a few hours. 

\item The \feka\ line, the shape of the optical/UV continuum, and the
Eddington ratios in the three quasars suggest the presence of a
standard accretion disk in the core. From the observed EW of a broad
\feka\ line of $\sim$ 70~eV, we infer that the jet contributes 50\% of
the total flux in 2--10~keV. 

\item The SEDs of the three sources are well described by a model     
including both the jet and the disk contributions. The latter
dominates the optical to soft X-ray continuum, while beamed emission
is responsible for the radio-to-IR flux and emission at energies $>$
2~keV. 

\item The jet kinetic power and disk luminosity follow the same trend
observed for other powerful blazars, implying that a large fraction of
the accretion power is converted into bulk kinetic energy of the jet. 
   
\end{itemize}      

Future coordinated observations, especially at gamma-rays with \glast,
will be necessary to confirm the estimates of the jet power and
efficiency. Indeed, in the most powerful blazars the bulk of the total
luminosity emerges at GeV energies, making this band of crucial
importance for the study of jets. 

\acknowledgements     
     
This project is funded by NASA grants NAG-30240 and HST-GO4-5111A, and
LTSA grant NAG5--10708 (MG). FT and LM acknowledge support from
COFIN/MIUR grant 2004023189-005.


\clearpage     

\begin{center}
\begin{tabular}{llcrlrr}
\multicolumn{7}{l}{{\bf Table 1: The Targets}} \\
\multicolumn{7}{l}{   } \\ \hline
& & & & & & \\
Source & $z$ & Gal N$_H$ & Class. & m$_V$ & log L$_{BLR}$ & M$_{BH}$\\
& & & & & & \\
(1) & (2) & (3) & (4) & (5) & (6) & (7) \\
& & & & & & \\ \hline
& & & & & & \\
0723+679 & 0.846 & 4.31 & SSRQ &  18.0 & 44.76 & 4.6 \\
1136--135& 0.554 & 3.5& FSRQ & 16.1 & 45.17 & 4.6 \\
1150+497 & 0.334 & 2.0 & FSRQ & 17.1 & 44.36 & 3.3 \\
& & & & & & \\ \hline

\end{tabular}
\end{center}

\noindent
{\bf Explanation of Columns:} 1=Source IAU name; 2=Redshift;
3=Galactic column density in $10^{20}$ \nh; 4=Radio Classification:
Steep Spectrum Radio Quasar (SSRQ), Flat Spectrum Radio Quasar (FSRQ);
5=Core optical V magnitude; 6=Total luminosity of the BLR (Cao \&
Jiang 1999); 7=Black hole mass in $10^8$ M$_{odot}$, from Woo \& Urry (2002);
Oshlack, Webster, \& Whiting (2002); Shields et al. (2003).

     
\clearpage     

\begin{center}
\begin{tabular}{ll}
\multicolumn{2}{l}{{\bf Table 2: Best-fit of X-ray Continua}} \\
\multicolumn{2}{l}{   } \\ \hline
& \\
Source & Best-fit Model and Parameters \\
& \\
& \\ \hline
& \\
0723+679 & Broken Power Law \\
               & $\Gamma_{soft}$= 1.84 $\pm$ 0.04 \\
               & $\Gamma_{hard}$= 1.65$^{+0.07}_{-0.10}$ \\
               & E$_{break}=1.89^{+1.45}_{-0.47}$ keV \\ 
               & $\chi^2_r$=0.98/602 \\
               & F$_{2-10~keV}=8 \times 10^{-13}$ \flux \\
               & L$_{2-10~keV}=2 \times 10^{45}$ \lum \\
               & \\ 
1136--135& Broken Power Law \\
               & $\Gamma_{soft}$= 1.91 $\pm$ 0.06 \\
               & $\Gamma_{hard}$= 1.72 $\pm$ 0.06 \\
               & E$_{break}=1.55 \pm 0.25$ keV \\ 
               & $\chi^2_r$=1.03/257 \\
               & F$_{2-10~keV}=1.3 \times 10^{-12}$ \flux \\
               & L$_{2-10~keV}=8.4 \times 10^{44}$ \lum \\
               & \\ 
1150+497 & Broken Power Law \\
               & $\Gamma_{soft}$= 2.10 $\pm$ 0.08 \\
               & $\Gamma_{hard}$= 1.72 $\pm$ 0.05 \\
               & E$_{break}=1.85 \pm 0.35$ keV \\ 
               & $\chi^2_r$=1.07/324 \\
               & F$_{2-10~keV}=2.7 \times 10^{-12}$ \flux \\
               & L$_{2-10~keV}=6.0 \times 10^{44}$ \lum \\
               & \\ \hline
\end{tabular}
\end{center}

\noindent
{\bf Explanation of Columns:} 1=Source; 2=Description of parameters:
for the continuum, $\Gamma_{soft}$ and $\Gamma_{hard}$ are the photon
indices below and above the break energy, respectively; F$_{2-10~keV}$
and L$_{2-10~keV}$ the intrinsic (absorption-corrected) flux and
luminosity in 2--10 keV. 


\clearpage     

\begin{center}
\begin{tabular}{ll}
\multicolumn{2}{l}{{\bf Table 3: Fe K line}} \\
\multicolumn{2}{l}{   } \\ \hline
1136--135 & \underline{Narrow Gaussian} \\
                & E$_{rest}=6.32^{+0.06}_{-0.07}$ keV \\ 
                & $\sigma_{rest} \equiv 0.01$ keV \\
                & Flux=$2.8^{+1.7}_{-1.5} \times 10^{-6}$ ph cm$^{-2}$ s$^{-1}$ \\
                & $\chi^2_r$/dofs=1.01/254 \\
                & EW$_{rest}=62^{+34}_{-37}$ eV \\
                & \\                 
                & \underline{Broad Gaussian} \\
                & E$_{rest}=6.34^{+1.03}_{-2.06}$ keV \\ 
                & $\sigma_{rest} \equiv 0.25$ keV \\
                & $\chi^2_r$/dofs=1.06/254 \\
                & Flux=$1.9 (< 3.3) \times 10^{-6}$ ph cm$^{-2}$ s$^{-1}$ \\
                & EW$_{rest}=30 (<108)$ eV \\
                & \\          
1150+497 & \underline{Narrow Gaussian} \\
                & E$_{rest}=6.34^{+0.08}_{-0.06}$ keV \\ 
                & $\sigma_{rest}=0.01$ keV \\
                & $\chi^2_r$/dofs=0.89/232 \\
                & Flux=$3.8^{+2.4}_{-2.6} \times 10^{-6}$ ph cm$^{-2}$ s$^{-1}$ \\
                & EW$_{rest}=68^{+43}_{-38}$ eV \\
                & \\
                & \underline{Broad Gaussian} \\
                & E$_{rest}=6.58^{+0.46}_{-0.20}$ keV \\ 
                & $\sigma_{rest}=0.25$ keV \\
                & $\chi^2_r$/dofs=0.88/232 \\
                & Flux=$5.9^{+5.6}_{-4.7} \times 10^{-6}$ ph cm$^{-2}$ s$^{-1}$ \\
                & EW$_{rest}=81^{+77}_{-65}$ eV \\
                & \\ \hline 
                & \\
\end{tabular}
\end{center}

\noindent
{\bf Explanation of Columns:} 1=Source; 2=Description of parameters:
E$_{rest}$ and $\sigma_{rest}$ are the Gaussian center energy and width in the source
rest-frame, EW the line Equivalent Width, Flux is the line flux. 


\clearpage     


\begin{center}
\begin{tabular}{rlrllllrrr}
\multicolumn{10}{l}{{\bf Table 4: SEDs Parameters$^{a}$}} \\
\multicolumn{10}{l}{   } \\ \hline
& & & & & & & & \\
Source & $\delta$ & $\gamma_b$ & $\gamma_2$ & $K$ & $B$ & R$_{BLR}$ & L$_{disk}$ & L$_{jet}$ & P$_{jet}$ \\
& & & & & & & & & \\
(1) & (2) & (3) & (4) & (5) & (6) & (7) & (8) & (9) & (10) \\
& & & & & & & & & \\ \hline
& & & & & & & & & \\
0723+679 & 15 & 100 & 1 & 7   & 2 & 2.0  & 10 & 0.8 & 350 \\
1136--135& 15 & 150 & 1 & 1.5 & 2 & 5.2  & 5  & 0.4 & 73  \\
1150+497 & 20 &  30 & 3 & 1   & 4 & 2.6  & 3  & 0.1 & 76  \\
& & & & & & & & & \\ \hline

\end{tabular}
\end{center}

\noindent
\noindent{\bf Note (see also column explanations):} 
a=Additional parameters: low energy boundary $\gamma_1=1$, broken power law  
spectral indices $n_1$=2, $n_2$=3.5, emission region radius $R=10^{16}$ cm. 

{\bf Explanation of Columns:} 1=Source IAU name; 
2=Doppler factor; 
3=Break energy of the broken-power-law distribution of particles with spectral indices $n_1,2$ below and above the break (see Note); 
4=High energy boundary of the broken power law (in 10$^4$); 
5=Normalization of the particle distribution (in 10$^4$ cm$^{-2}$ s$^{-1}$); 
6=Magnetic field (in Gauss); 
7=Size of the BLR (in $10^{17}$ cm); 
8=Disk luminosity (in $10^{45}$ \lum); 
9=Jet luminosity (in $10^{45}$ \lum); 
10=Jet power (in $10^{45}$ \lum). 

     
\clearpage
     
\begin{figure}     
\centerline{\includegraphics[width=7cm,height=6cm]{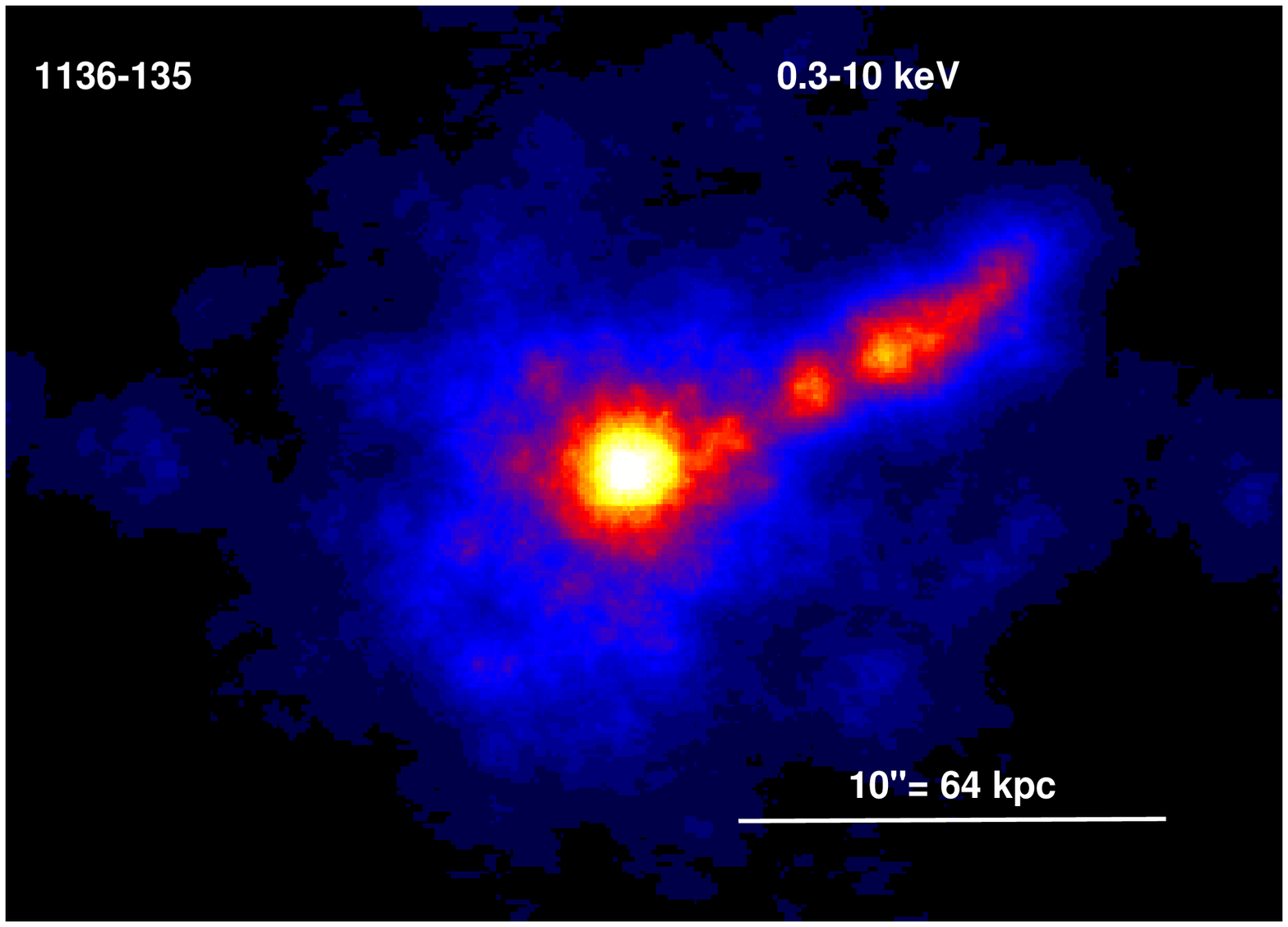}}     
\centerline{\includegraphics[width=7cm,height=6cm]{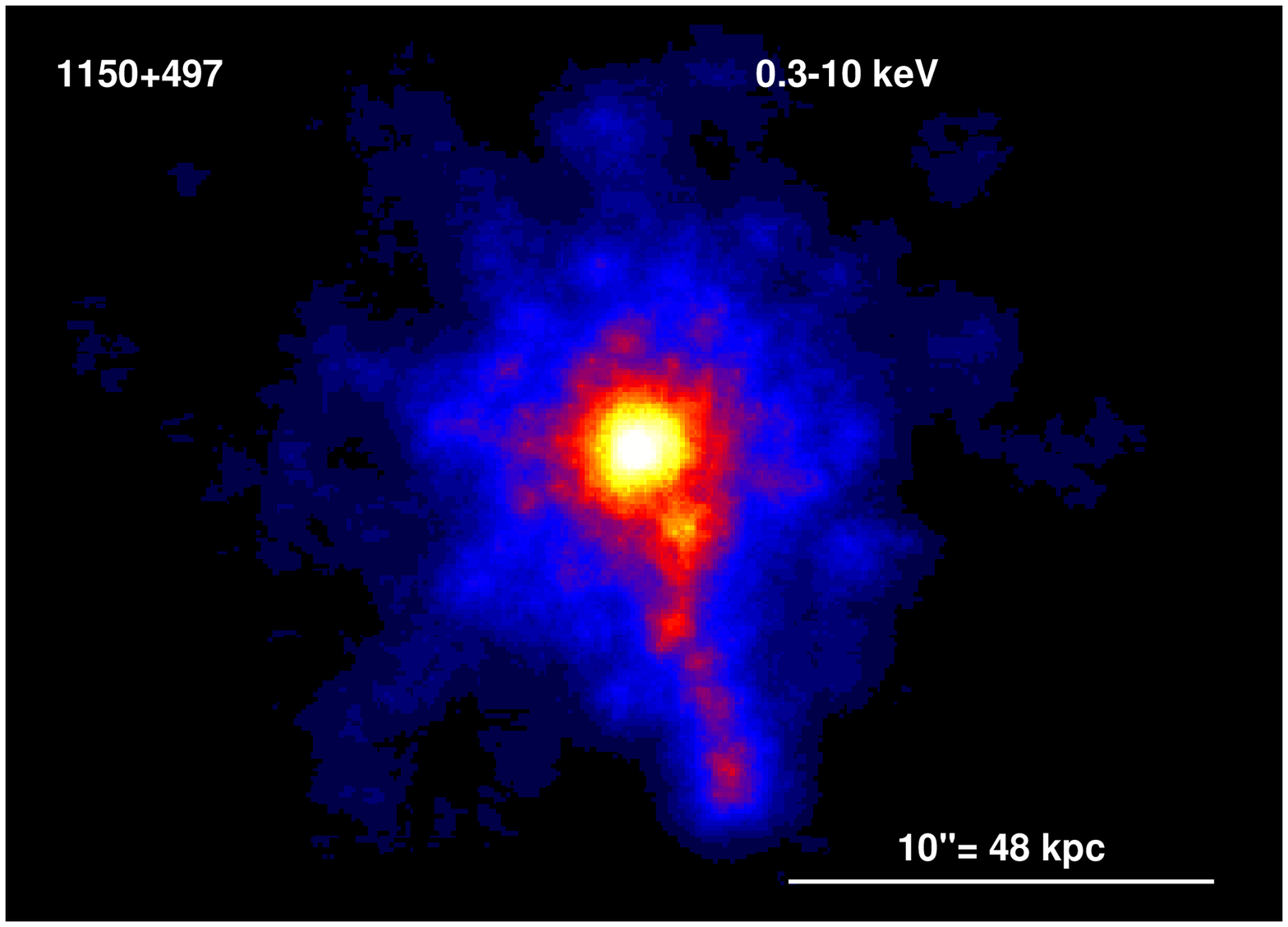}}     
\centerline{\includegraphics[width=7cm,height=6cm]{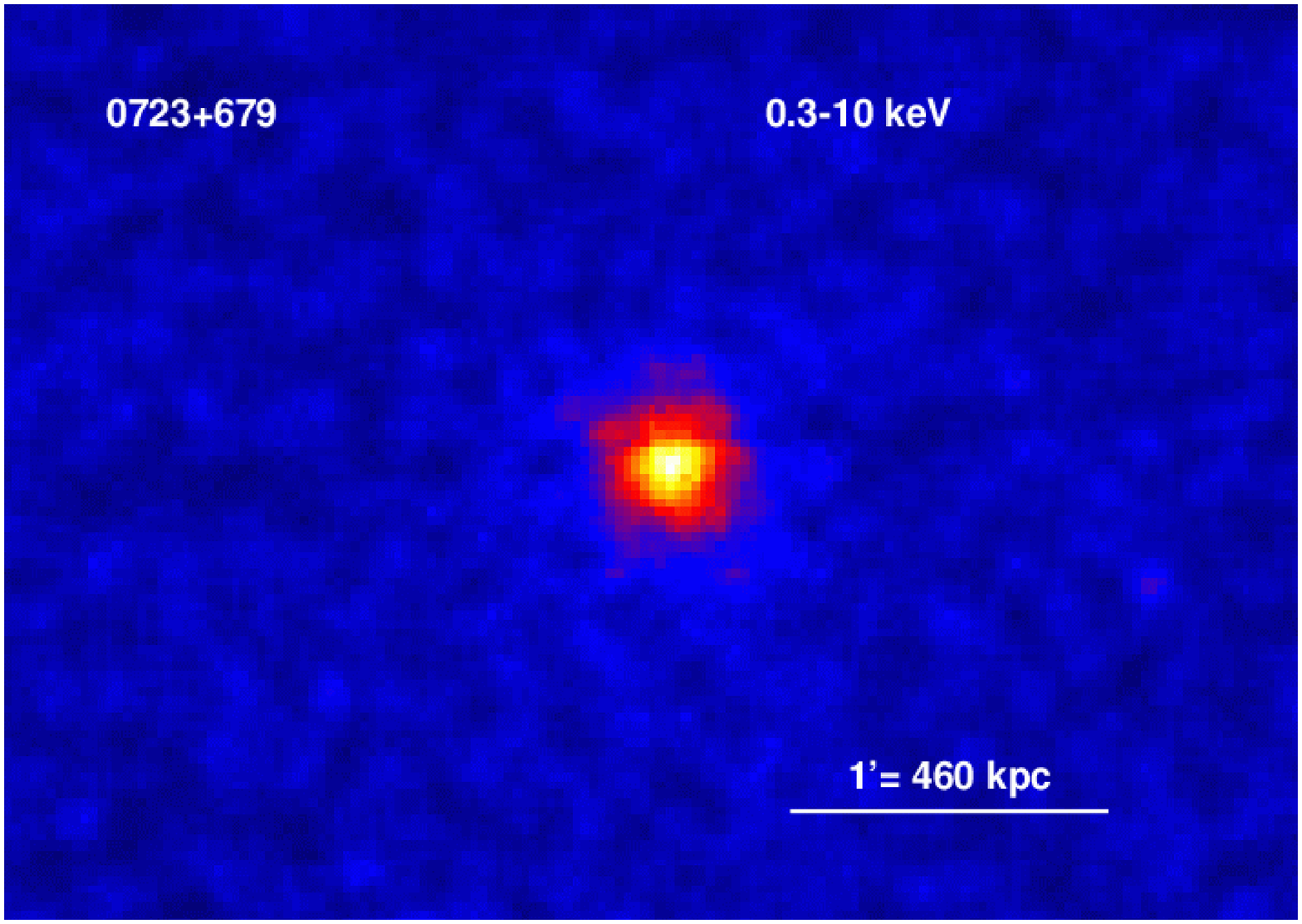}}     
\caption{ACIS-S and EPIC MOS1 images of the three sources.      
The ACIS images were smoothed using the sub-package {\it fadapt} of
{\it FTOOLS} with a circular top hat filter of adaptive size in order
to achieve a minimal number of 10 counts under the filter; each final
pixel is 0.1\arcsec. The MOS1 data were rebinned by a factor 32 and
smoothed the same way as the ACIS images but with minimal number of
counts 30; each final pixel is 1.6\arcsec. North is up and East to the
left.} 
\end{figure}


\begin{figure}     
\centerline{\includegraphics[height=9cm,width=10cm]{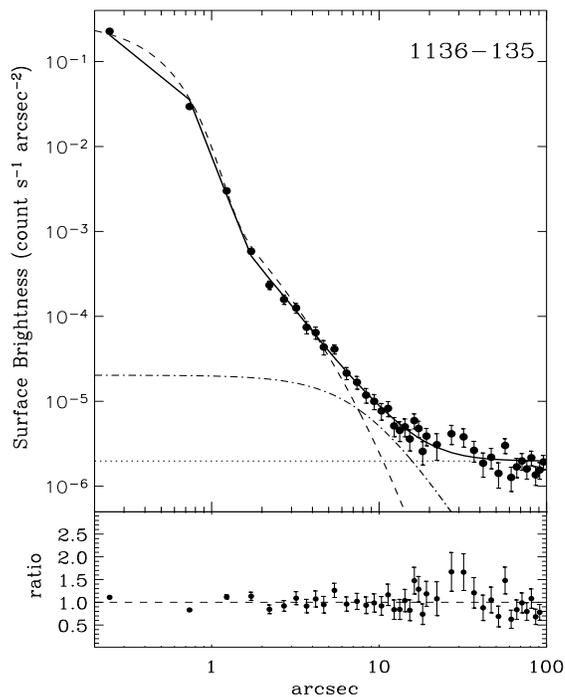}}     
\vspace{0.5in}\centerline{\includegraphics[height=9cm,width=10cm]{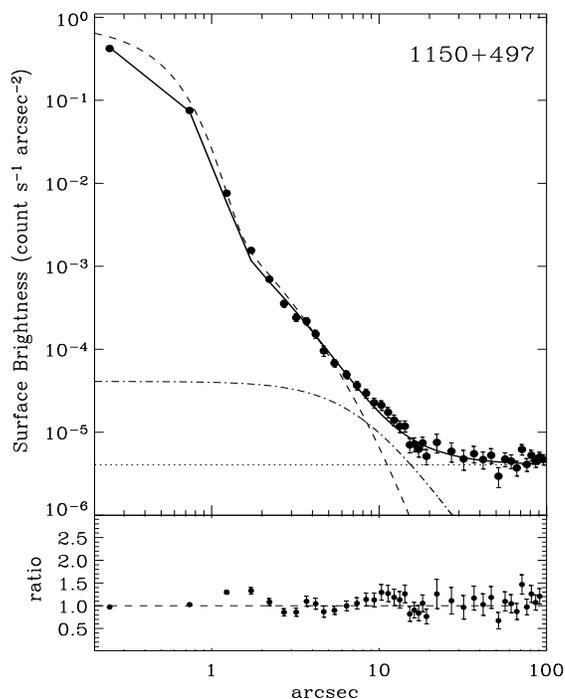}}     
\caption{X-ray surface brightness profiles of 1136--135 and 1150+497. 
The solid line represents the best-fit model, which comprises     
the PSF model (dashed line), $\beta$--model (dot-dashed line), and     
the background level (dotted line).     
The lower panels show the data-to-model ratios. 
}
\end{figure}     
     
\begin{figure}     
\includegraphics[width=13cm,height=8cm]{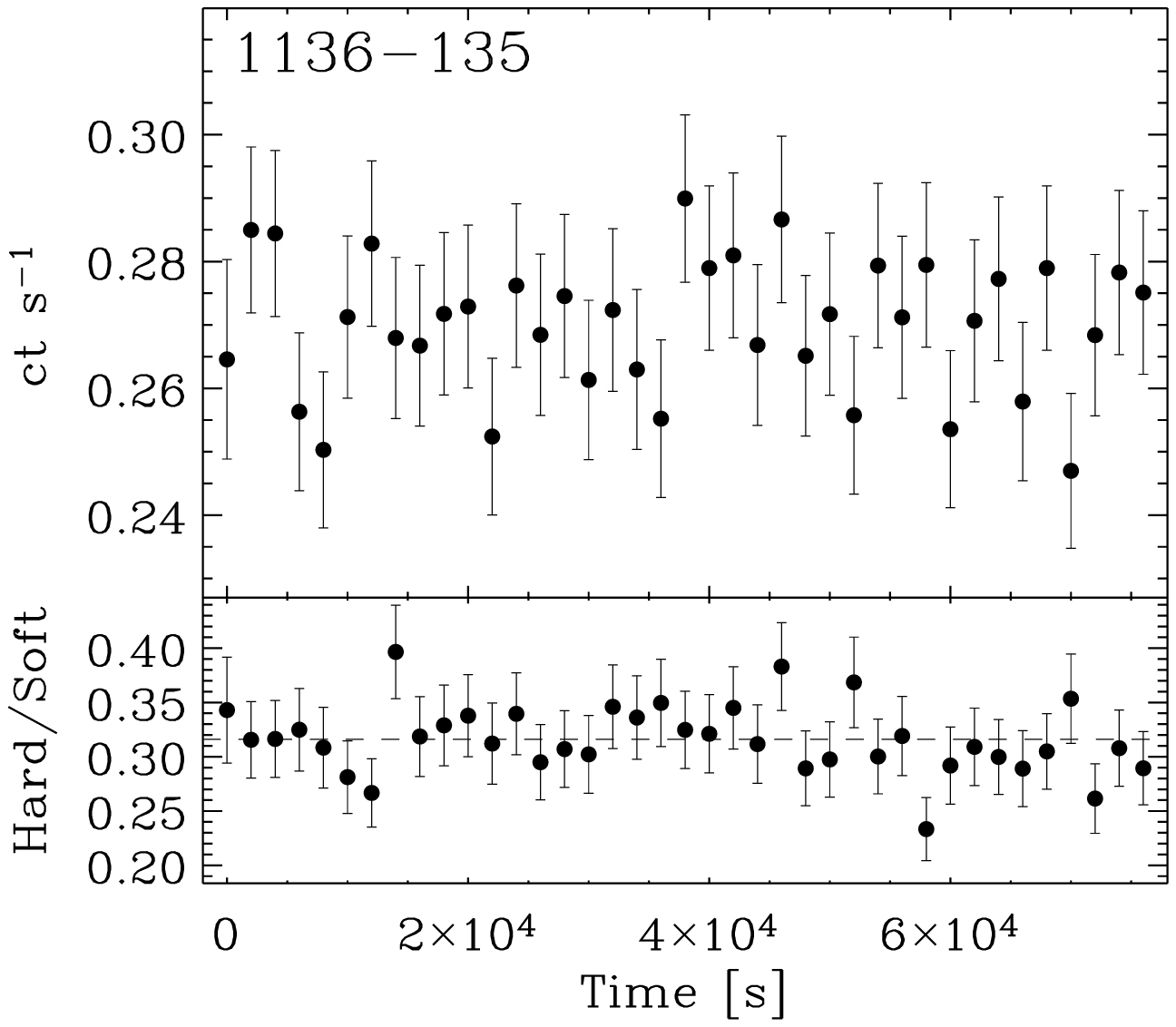}
\vspace{0.8in}     
\includegraphics[width=13cm,height=8cm]{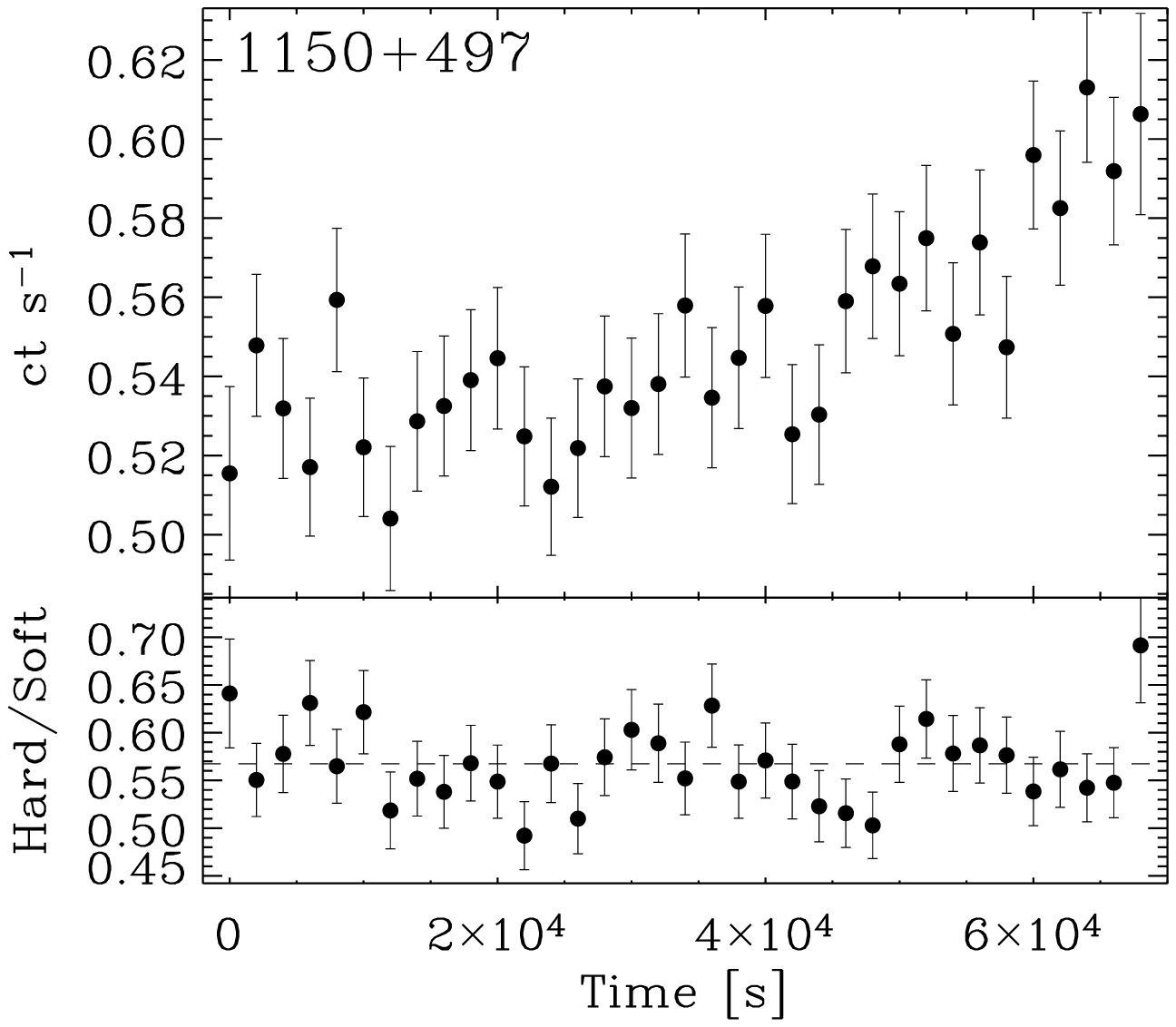}
\vspace{0.5in}     
\caption{Background-subtracted ACIS-S light curves 
in the 0.3--10 keV energy range. 
}
\end{figure}     
     
\begin{figure}
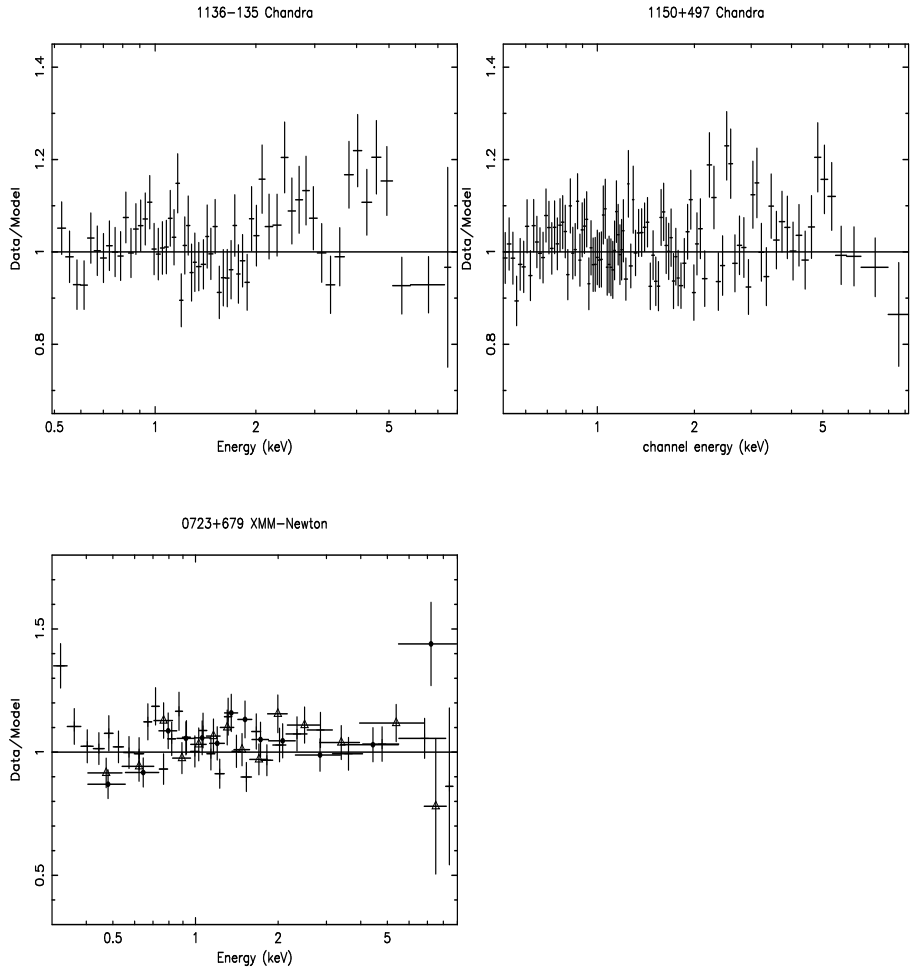
     
\includegraphics[height=6cm,width=6cm,angle=-90]{f4a.ps}\includegraphics[height=6cm,width=6cm,angle=-90]{f4b.ps}

\vspace{0.3in}
\includegraphics[height=6cm,width=6cm,angle=-90]{f4c.ps}
\caption{Residuals of the best-fit continuum model (broken power law     
and Galactic absorption) to the ACIS-S and EPIC spectra of the three     
sources. For the EPIC data: {\it crosses}, pn; {\it filled dots},     
MOS1; {\it open triangles,} MOS2. An Fe-K line is visible in the     
\chandra\ residuals of 1136--135 and 1150+497 around 4--5~keV. 
}     
\end{figure}     
     
\clearpage
     
     
     
     
     
\begin{figure}     
\centerline{\includegraphics[height=15cm,width=13cm]{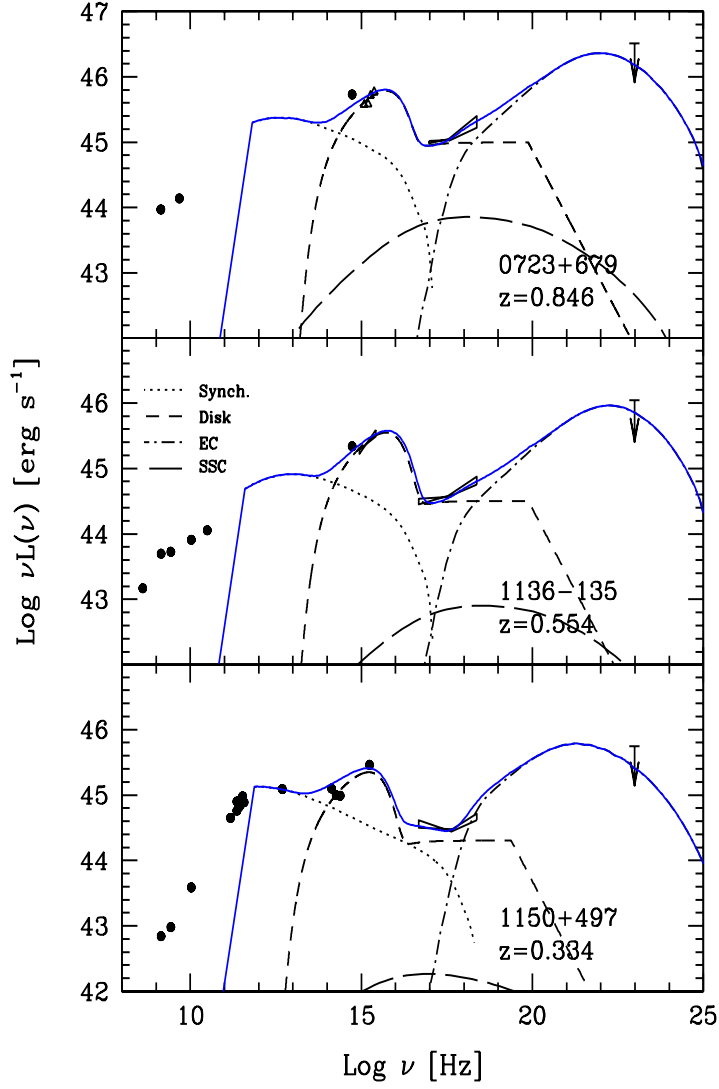}}     
\caption{Spectral Energy Distributions of the cores of the three     
quasars (see text). The lines represent the models used to fit the     
data. {\it Solid line:} Total spectrum;      
{\it Dotted line:} synchrotron;      
{\it Long dashed line:} synchrotron-self Compton (SSC); 
{\it Dot-dashed line:} External Compton (EC);      
{\it Short dashed line:} Disk emission. }     
\end{figure}

     
\begin{figure}     
\centerline{\includegraphics[height=15cm,width=13cm]{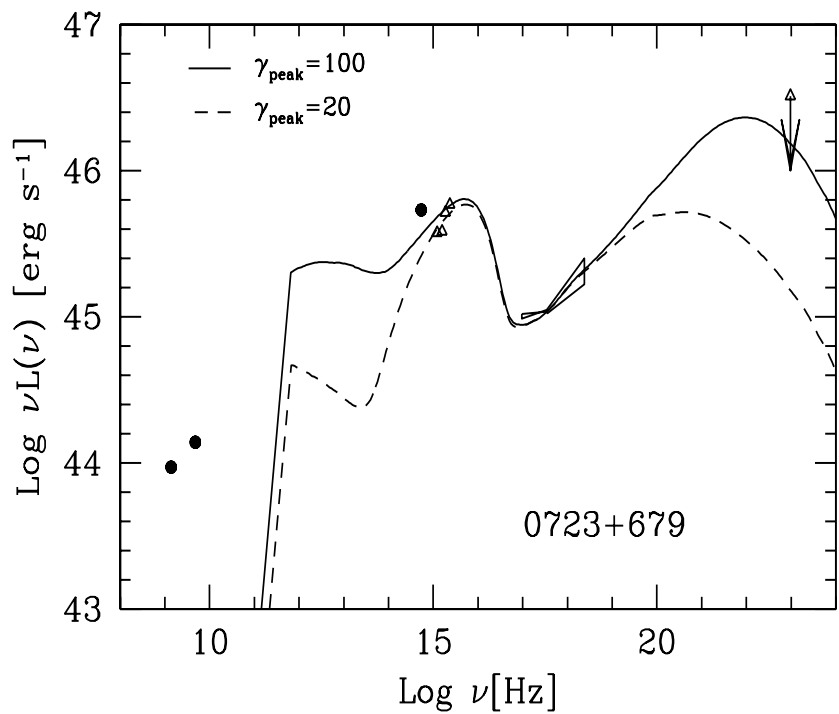}}     
\caption{Spectral Energy Distribution of 0723+679 for two different     
values of the electron distribution break energy. For simplicity, we     
only show the sum of the various model components (dashed and solid     
line). The GeV energy band is crucial to constrain the value of     
$\gamma_{peak}$. 
}     
\end{figure}

     
\clearpage     
     
\begin{figure}     
\centerline{\includegraphics[height=15cm,width=13cm]{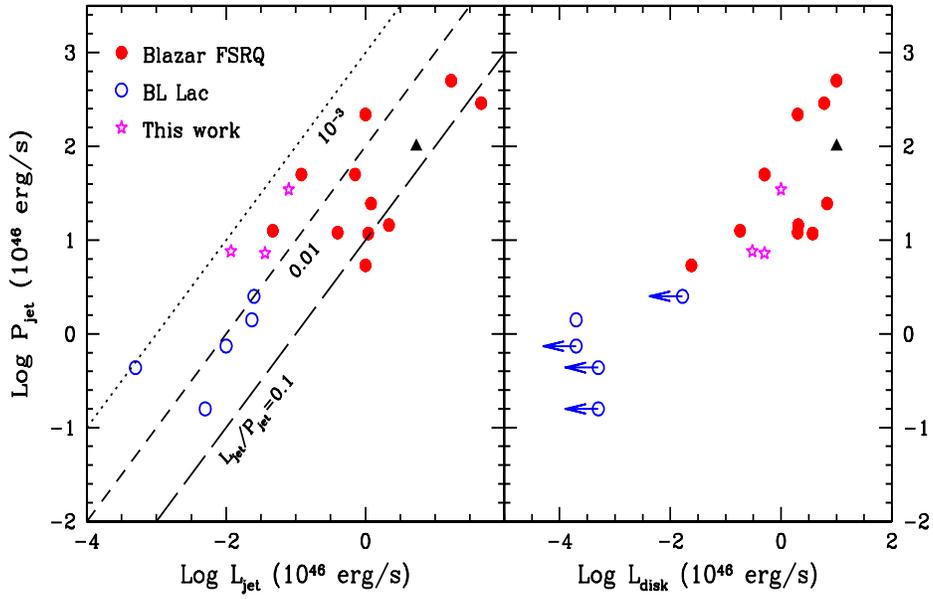}}     
\caption{Comparison of the jet power P$_{jet}$, 
calculated with the parameters in Table~4, with the jet luminosity
{\it (left panel)} and the disk luminosity {\it (right panel)}. The
FSRQs and BL Lacs data are from Maraschi \& Tavecchio (2003). The
black triangle is the $z=2.979$ blazar SWIFT~J0746.3+2548 (Sambruna et
al. 2006a).}
\end{figure}

     
\end{document}